\documentclass[aps,preprint, groupedaddress, floatfix, nofootinbib]{revtex4-1}

\usepackage[english]{babel}

\usepackage[letterpaper,top=2cm,bottom=2cm,left=3cm,right=3cm,marginparwidth=1.75cm]{geometry}

\usepackage{amsmath}
\usepackage{graphicx}
\usepackage[colorlinks=true, allcolors=blue]{hyperref}
\usepackage[toc,page,titletoc]{appendix}
\usepackage{amssymb}

\usepackage{epsfig}
\usepackage{color}
\begin{document}

\title{Hubbard-$U$-corrected electron-phonon interactions in strongly correlated materials via the finite-displacement method}

\author{Jiale Chen$^{1,2,3}$\footnote{These authors contributed equally to this work.}, Youyou Tu$^{4}$\footnotemark[1], Chengliang Xia$^{2}$, Jin Zhao$^{4,5,6}$\footnote{Correspondence to zhaojin@ustc.edu.cn or hanghui.chen@nyu.edu.} and Hanghui Chen$^{2,3,7}$\footnotemark[2]}

\affiliation{$^1$Key Laboratory of Polar Materials and Devices, Ministry of Education, East China Normal University, Shanghai 200241, China\\
  $^2$NYU-ECNU Institute of Physics, NYU Shanghai, Shanghai 200124, China\\
  $^3$School of Information and Electronic Engineering, East China Normal University, Shanghai 200241, China\\
  $^4$Department of Physics and ICQD/Hefei National Research Center for Physical Sciences at the Microscale, University of Science and Technology of China, Hefei, Anhui 230026, China\\
  $^5$Department of Physics and Astronomy, University of Pittsburgh, Pittsburgh, Pennsylvania 15260, USA\\
  $^6$Synergetic Innovation Center of Quantum Information \& Quantum Physics, University of Science and Technology of China, Hefei, Anhui 230026, China\\
  $^7$Department of Physics, New York University, New York, NY 10003, USA
}

\begin{abstract}
The interplay between electron-electron and electron-phonon
interactions is one of the central topics in condensed matter physics.
Although the density functional theory plus Hubbard $U$ correction method
(DFT+$U$) is broadly used to study electronic structure of
strongly correlated materials, the extension of this method to
electron-phonon $g$ matrices and electron-phonon coupling has received
limited attention. In this work, we implement an algorithm that
integrates DFT+$U$ method with the finite-displacement method for the
calculations of phonons and electron-phonon $g$ matrices. The Hubbard
$U$ corrections are applied not only to electronic and phonon
structures, but, more importantly, also to electron-phonon $g$
matrices. We demonstrate our algorithm in two prototypical correlated
materials: infinite-layer nickelates LaNiO$_2$ and ruthenium dioxide
RuO$_2$. We find that: i) While the Hubbard
$U$ corrections weakly increase the electron-phonon interaction of
20\% hole-doped LaNiO$_2$, its total electron-phonon coupling remains
small and is insufficient to account for the observed
superconducting transition temperature of about 10-30 K. Our results
contrast with the recent work~\cite{stevein_twogap} showing that the
full GW corrections yield an elevated electron-phonon coupling of 20\%
hole-doped LaNiO$_2$ five times larger than its DFT value.
We attribute this discrepancy to the differences in the
Fermi surface topology between DFT+$U$ and GW methods.
ii) The inclusion of Hubbard $U$ corrections eliminates
  the imaginary phonon modes of RuO$_2$ under strain on the TiO$_2$
  substrate and substantially reduces the electron-phonon
  coupling. Our results alleviate the discrepancy between the
previously reported large theoretical electron-phonon coupling and the
low superconducting transition temperature observed
experimentally. Our work provides an algorithm that fully includes the
Hubbard $U$ corrections on electron-phonon properties of correlated
materials, and highlights the importance of Fermi surface shape and
correlation effects on phonon spectrum and electron-phonon $g$ matrices.
\end{abstract}

\maketitle

\section{Introduction}
The interplay between electron-electron and electron-phonon
interactions has been one of the most important and active research
areas in condensed matter
physics~\cite{KULIC20001,PhysRevB.78.193404,PhysRevB.92.085132,Preempted,stevein_twogap,PhysRevLett.126.146401,PhysRevLett.125.147001,Strain-stabilized,PhysRevLett.133.186501}. 
In most calculations of electron-phonon properties of real materials, the
electron-phonon interactions and electron-electron interactions are
treated separately~\cite{Preempted,Strain-stabilized}. The total
electron-phonon spectrum $\alpha^2F(\omega)$ and electron-phonon
coupling $\lambda$ are obtained from plain density functional theory
(DFT) calculations, while the electron-electron interactions are
modelled via an empirical Anderson-Morel parameter
$\mu^{*}$~\cite{PhysRev.125.1263}, whose value is usually between 0.1
and 0.2~\cite{PhysRevLett.125.147001,Strain-stabilized,PhysRevB.66.020513,Nat.Com.2021.2314,Nat.Com.2024.7.33,PhysRevB.101.060506,PhysRevB.54.16487}.
Both $\lambda$ and $\mu^{*}$ enter the McMillan-Allen-Dynes
formula~\cite{Tc1}, which is used to estimate the superconducting
transition temperature. However, in strongly correlated materials, the
electron-electron and electron-phonon interactions are
intertwined~\cite{PhysRevLett.126.146401,stevein_twogap,Npj.com.mat.3.2057,PhysRevLett.133.186501}.
Therefore the correlation effects on the electron-phonon interactions
need to be modelled in a more direct manner. Recently an approach that
includes full GW corrections~\cite{stevein_twogap} in the calculations
of electron-phonon interactions has been developed and has been
successfully applied to prototypical strongly correlated materials
such as cuprates and nickelates. In such an approach, the GW
corrections are applied not only to the electronic structure, but also
to the phonon spectrum and electron-phonon $g$ matrices.  However, the
computational cost of full GW corrections into electron-phonon
interaction calculations remains prohibitively high, limiting their
application to complex real materials. On the other hand, DFT+$U$
method has been widely used to calculate the electronic structure of
strongly correlated materials, due to its computational
efficiency~\cite{PhysRevLett.133.186501,PhysRevLett.131.126001,PhysRevX.10.011024,PhysRevB.44.943,PhysRevB.49.14211,PhysRevX.11.041009,PhysRevB.44.943,npj.qtm.mat.1.2397,Npj.qtm.mat.31.2397,Jou.phy.4.767.0953}.
Nevertheless, the impact of Hubbard $U$ corrections on phonon spectrum
and electron-phonon $g$ matrices has not been systematically
investigated.

In this work, we implement an algorithm that combines the DFT+$U$ and finite displacement methods~\cite{TOGO20151,PhysRevLett.78.4063,arXiv.2508.14852,arXiv.2511.21905} to calculate phonon spectrum and electron-phonon $g$ matrices, which enables to calculate the fully Hubbard $U$ corrected electron-phonon interactions of correlated materials. In our approach, the Hubbard $U$ corrections are not only applied to electronic structure, but also to phonon spectrum and electron-phonon $g$ matrices. We use our approach to study two representative correlated materials: infinite-layer nickelates (20\% hole-doped LaNiO$_2$) and ruthenium dioxides (RuO$_2$).

Infinite-layer nickelate~\cite{nc,Hepting2020,PhysRevX.10.011024,PhysRevB.70.165109,PhysRevB.100.201106,PhysRevX.10.021061,PhysRevX.11.011050,PhysRevB.105.115134,17orb,PhysRevLett.125.027001,Gu2020,Wang2021,PhysRevLett.125.147003,PhysRevLett.125.027001,PhysRevLett.125.147003,Lee2023,Wang2021,Gu2020,doi:10.1021/acs.nanolett.0c01392,PhysRevMaterials.4.121801,Ad.Mat.2104083.2021,doi:10.1126/sciadv.abl9927,Adv.Mat.6.341.2023,10.3389/fphy.2022.834658,84jh-sx4m,xqm6-wr7n} has sparked tremendous interest, owing to its superconducting properties. The plain DFT calculations show that the electron-phonon coupling of infinite-layer nickelate is small and thus electron-phonon interactions alone are not sufficient to account for its superconductivity~\cite{PhysRevB.108.174512,PhysRevB.107.075159,PhysRevB.100.205138}. However, a recent study shows that the full GW corrections to the electron-phonon coupling of 20\% hole-doped LaNiO$_2$ result in a five times increase over the plain DFT value~\cite{stevein_twogap}. We find that with the full Hubbard $U$ corrections, the electron-phonon interactions in 20\% hole-doped LaNiO$_2$ still remains small in a wide and physically reasonable range of Hubbard $U$ applied on Ni-$d$ orbitals. We attribute this marked discrepancy between DFT+$U$ and GW methods to the differences in the Fermi surface topology obtained from the two approaches.  

Ruthenium dioxide RuO$_2$~\cite{Nat.com.1.3784.2041,PhysRevResearch.3.033214,Phy.che.B.40.12677} has also drawn great attention because it is one of the candidates for altermagnetism~\cite{NatCommun.16.63344,NatCommun.16.60891,JMaterChemC.13,NanoConvergence.13.1,Nat.elc.735.2520.5,PhysRevX.12.040501,HUSSAIN2025417723,6fxv-153y,PhysRevLett.118.077201}. But currently it is under debates~\cite{npjSpintronics.2.55,Kiefer_2025,YUMNAM2025102852}. When RuO$_2$ is grown in a thin film form on (110)-oriented TiO$_2$ substrates it exhibits superconductivity with a transition temperature of about 1.5 K~\cite{PhysRevLett.133.176401,PhysRevLett.125.147001,Strain-stabilized}. While the experimental transport measurement of TiO$_2$-strained RuO$_2$ is compatible with conventional superconductivity, the plain DFT calculations predict imaginary phonon modes in the experimentally observed orthorhombic $Pnnm$ crystal structure, along with an unphysically large electron-phonon coupling~\cite{PhysRevLett.125.147001,Strain-stabilized}. We find that the inclusion of Hubbard $U$ corrections on Ru-$d$ orbitals eliminates the imaginary phonon modes and dynamically stabilizes RuO$_2$ thin films in the experimentally observed crystal structure. Furthermore, increasing Hubbard $U_{\rm{Ru}}$ substantially reduces the total electron-phonon coupling, which helps explain the low superconducting transition temperature observed experimentally. 

The algorithm is explained in the Result section and the complete computational details are provided in the Methods section.

\section{Results}


\subsection{Finite-displacement method with Hubbard $U$ correction}

The most widely-used \textit{ab initio} method for computing electron-phonon
properties of real solids is density functional perturbation theory (DFPT)~\cite{PhysRevLett.133.176401,RevModPhys.73.515}---a
perturbative approach within Kohn-Sham DFT framework, combined with
Wannier function interpolation~\cite{PhysRevB.76.165108}. In this
approach, DFT and DFPT calculations on a coarse Brillouin zone grid
provide electronic structure and phonon perturbation
potential, which serve as the inputs for subsequent interpolation~\cite{npj.Com.Mat.156.2057,EPW,Perturbo}.

To incorporate electron correlation effects in electron-phonon
properties, the density functional perturbation theory plus Hubbard
$U$ corrections (DFPT+$U$) method has been developed for correlated
electron systems~\cite{PhysRevLett.127.126404}. However, DFPT+$U$
sometimes encounters convergence issues. For example, in our study of
infinite-layer nickelates, calculations using Quantum Espresso
(QE)~\cite{QE} based on the DFPT+$U$ method fail to
produce converged phonon results when Hubbard $U_{\rm{Ni}}$ exceeds 2 eV.

To overcome this challenge, we instead utilize the finite-displacement
method~\cite{TOGO20151,PhysRevLett.78.4063,arXiv.2508.14852,arXiv.2511.21905}, which circumvents the convergence issues of DFPT+$U$. The
finite-displacement method has been successfully applied to compute
electron-phonon properties of various solid
materials~\cite{PhysRevB.85.115317,PhysRevB.93.035414,Monserrat_2018,PhysRevB.101.184302,PhysRevB.106.094316}. It
involves performing a set of supercell DFT calculations to obtain the
phonon perturbation potential in real space. The electron-phonon $g$ matrix
in momentum space is then derived via Fourier transform. Crucially,
as long as the individual supercell DFT+$U$ calculations are feasible,
this method yields electron-phonon results that fully incorporate the
Hubbard $U$ corrections.


In this work, we implement the finite-displacement method using DFT
software based on numeric atomic orbital (NAO) basis sets. Compared
to the usual plane-wave bases, the spatial localization of NAO bases benefit
NAO-based DFT software with superior computational efficiency for
large-scale simulations, such as the supercell calculations required
here. The electron-phonon coupling matrix element $g_{mn}^\nu(\mathbf{k},\mathbf{q})$ is
defined as:
\begin{equation}
g_{mn}^\nu(\mathbf{k},\mathbf{q})
=\langle\psi_{m\mathbf{k}+\mathbf{q}}|
\partial_{\mathbf{q}\nu}V
|\psi_{n\mathbf{k}}\rangle\label{r1}
\end{equation}
It couples the initial Bloch state $|\psi_{n\mathbf{k}}\rangle$ to the final state $|\psi_{m\mathbf{k}+\mathbf{q}}\rangle$ through the $\nu$th branch phonon with crystal momentum $\mathbf{q}$. The phonon-induced perturbation potential $\partial_{\mathbf{q}\nu}V$ can be expanded using derivatives of the potential with respect to atomic displacements ${\partial V}/{\partial \tau_{l \lambda \textbf{R}_p}}$
\begin{equation}
\partial_{\mathbf{q}\nu}V
=\sum\limits_{\textbf{R}_p}\sum\limits_{l,\lambda}
\sqrt{\frac{\hbar}{2M_l\omega_{\mathbf{q}\nu}}}
e_{l\lambda}^{\mathbf{q}\nu}
\frac{\partial V}{\partial\tau_{l\lambda{\mathbf{R}_p}}}
e^{i\mathbf{q}\cdot\mathbf{R}_p}
\label{r2}
\end{equation}
where $\mathbf{R}_p$ labels a lattice vector for a periodic unit cell; $l$ labels an atom; $\lambda=x,y,z$ is a Cartesian direction index; $M_l$ is the atomic mass. $\omega_{\mathbf{q}\nu}$ and $e_{l\lambda}^{\mathbf{q}\nu}$ represent phonon frequency and polarization vector, respectively. Substitute Eq.~\eqref{r2} into Eq.~\eqref{r1} gives:
\begin{equation}
g_{mn}^\nu(\mathbf{k},\mathbf{q})
=\sum\limits_{\textbf{R}_p}\sum\limits_{l,\lambda}
\sqrt{\frac{\hbar}{2M_l\omega_{\mathbf{q}\nu}}}
e_{l\lambda}^{\mathbf{q}\nu}
\left\langle\psi_{m\mathbf{k}+\mathbf{q}}\left|
\frac{\partial V}{\partial\tau_{l \lambda {\mathbf{R}_p} }}
\right|\psi_{n\mathbf{k}}\right\rangle
e^{i\mathbf{q}\cdot\mathbf{R}_p}\label{r3}
\end{equation}
The Kohn-Sham orbitals are expanded in the NAO basis:
\begin{equation}
|\psi_{n\mathbf{k}}\rangle
=\frac{1}{\sqrt{N}}\sum\limits_{\textbf{R}_e}
\sum\limits_{j}\sum\limits_{\beta}
c^{j \beta}_{n\mathbf{k}}e^{i\mathbf{k}\cdot\mathbf{R}_e}
|\phi_{ j \beta {\mathbf{R}_e}}\rangle\label{r4}
\end{equation}
where $c^{j \beta}_{n\mathbf{k}}$ is the expansion coefficient of $|\phi_{ j \beta {\mathbf{R}_e}}\rangle$ that represents $\beta$th NAO at atomic site $j$ in the unit cell with the lattice vector $\mathbf{R}_e$. $N$ is the number of unit cells. Inserting Eq.~\eqref{r4} into Eq.~\eqref{r3} allows us to express the potential derivative matrix elements as:
\begin{equation}
\begin{split}
&\left \langle \psi_{m\mathbf{k}+\mathbf{q}} \left|
\frac{\partial V}{\partial\tau_{l \lambda \mathbf{R}_p}}
\right |\psi_{n\mathbf{k}}\right\rangle\\
&=\frac{1}{N}\sum\limits_{\textbf{R}_e, \textbf{R}'_e}
\sum\limits_{\alpha,\beta}
\sum\limits_{i,j}
\left(c^{i\alpha}_{m\mathbf{k}+\mathbf{q}}\right)^*  c^{j\beta}_{n\mathbf{k}}
e^{-i(\mathbf{k}+\mathbf{q})\cdot\mathbf{R}_e}
e^{i\mathbf{k}\cdot\mathbf{R}'_e}
\left\langle\phi_{ i\alpha {\mathbf{R}_e} }\left|
\frac{\partial V}{\partial\tau_{ l \lambda {\mathbf{R}_p} }}
\right|\phi_{j \beta {\mathbf{R}'_e} }\right\rangle\label{r5}
\end{split}
\end{equation}
In Eq.~\eqref{r3} and Eq.~\eqref{r5}, the phonon dispersion $\omega_{\mathbf{q}\nu}$ and eigenvectors $e_{l\lambda}^{\mathbf{q}\nu}$ are obtained from the finite-displacement method via supercell force calculations. The potential derivative matrix elements in the NAO basis $\langle\phi_{i \alpha {\mathbf{R}_e} }|{\partial V}/{\partial\tau_{l \lambda  {{\mathbf{R}}_p}}}|\phi_{j \beta {\mathbf{R}'_e}}\rangle$ are computed from finite differences of the Hamiltonian matrix for the displaced supercells. The band structure and expansion coefficient of NAO are obtained by diagonalizing the unit cell Hamiltonian.

Following this derivation, we apply DFT+$U$ consistently to all calculations: the unit cell for bands, the supercells for forces and potential derivatives. This ensures that electronic structure, phonon and perturbation potential all incorporate the same Hubbard $U$ corrections. This leads to a self-consistent and accurate Hubbard $U$ description of electron-phonon properties. We note that since this finite-displacement approach only involves electronic self-consistent calculations, it bypasses the phonon convergence issue that we encounter in the DFPT+$U$ method.

\subsection{Infinite-layer nickelates}


\begin{figure}[t]  
\centering  
\includegraphics[width=0.95\textwidth]{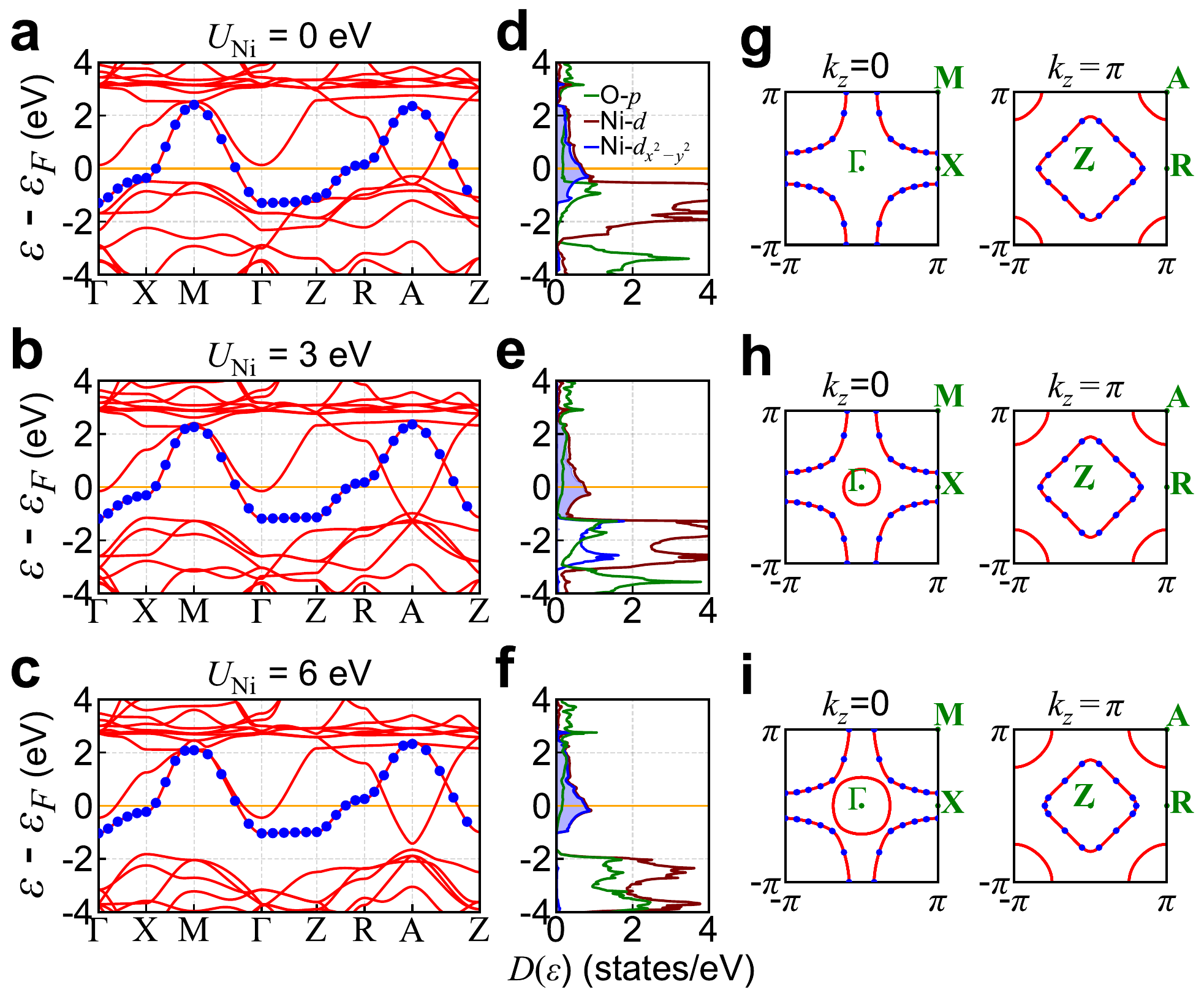}  
\caption{Electronic properties of 20\% hole-doped LaNiO$_2$: (a-c) Band structure calculated at $U_{\rm{Ni}}$ = 0, 3, 6 eV, respectively. The blue dots highlight that the band is derived from Ni-$d_{x^2-y^2}$ orbital. (d-f) Density of states calculated at $U_{\rm{Ni}}$ = 0, 3, 6 eV, respectively. The brown and green lines denote the projections onto Ni-$d$ and O-$p$ orbitals, respectively. The blue shaded region highlights the projection onto Ni-$d_{x^2-y^2}$ orbital. (g-i) Fermi surface calculated at $U_{\rm{Ni}}$ = 0, 3, 6 eV, respectively. Left sub-panel is for $k_z$ = 0 plane and right sub-panel is for $k_z$ = $\pi$ plane. The blue dots highlight that the Fermi sheet is derived from Ni-$d_{x^2-y^2}$ orbital.}
\label{fig:fig2}
\end{figure}

In this section, we study the fully Hubbard-$U$-corrected
electron-phonon interactions of infinite-layer nickelates: 20\%
hole-doped and undoped LaNiO$_2$. For clarity, we present the results of 20\%
hole-doped LaNiO$_2$ in the main text and the results of undoped LaNiO$_2$ in
the Supplementary Note 1 that includes Fig. S1-S5.
We find that doped and undoped infinite-layer
nickelates show similar results. For infinite-layer
nickelates, the onsite Hubbard $U$ interaction is applied to Ni-$d$
orbitals.

Figure~\ref{fig:fig2}(a-c) and (d-f) show the evolution of the
electronic band structure and the density of states (DOS) of 20\%
hole-doped LaNiO$_2$, calculated at a few representative Hubbard
$U_{\rm{Ni}}$. Since the total occupancy does not change, the
Ni-$d_{x^2-y^2}$ band (highlighted by the blue dots in (a-c) and the
blue shade in (d-f) remains pinned around the Fermi level and is
negligibly affected by the inclusion of Hubbard $U_{\rm{Ni}}$
corrections. In contrast, Hubbard $U_{\rm{Ni}}$ corrections lower the
energies of other Ni-$d$ bands and push them away from the Fermi
level, as shown by the purple curves in (d-f). We
  note that in Fig.~\ref{fig:fig2}(a-c), the relative flat bands
  around 3 eV are mainly derived from La-$4f$ orbitals. The partial
  density of states projected onto La-$4f$ orbitals, as well as the
  total density of state of 20\% hole-doped LaNiO$_2$ is shown in
  Fig.~S6 of the Supplementary Note 2. In addition, for all
$U_{\rm{Ni}}$ values in this study, there is a second band that also
crosses the Fermi level (known as the ``conduction band'')
~\cite{PhysRevB.100.201106}. As shown in (a-c), the conduction band is
pushed down in energy by Hubbard $U_{\rm{Ni}}$ corrections as well,
leading to a more pronounced ``self-doping'' effect at $\Gamma$ and A
points.  Fig.~\ref{fig:fig2}(g-i) display the corresponding Fermi
surface of 20\% hole-doped LaNiO$_2$, calculated at different
$U_{\rm{Ni}}$ values. Since the Fermi surface of 20\% hole-doped
LaNiO$_2$ is three-dimensional with a strong $k_z$ dependence, we show
Fermi surfaces at both $k_z=0$ and $k_z=\pi$ planes. We use blue dots
to highlight that the large Fermi sheet is derived from
Ni-$d_{x^2-y^2}$ orbital. At $k_z=0$ plane, we find that the
Ni-$d_{x^2-y^2}$ derived Fermi sheet is negligibly affected by Hubbard
$U_{\rm{Ni}}$ corrections. By contrast, Hubbard $U_{\rm{Ni}}$
corrections induce an additional $\Gamma$-centered electron pocket
that originates from the conduction band. At $k_z=\pi$ plane, Hubbard
$U_{\rm{Ni}}$ corrections have marginal effects, with the
Ni-$d_{x^2-y^2}$-derived Fermi sheet slightly shrinking and the
zone-corner pocket slightly expanding. Combining these observations,
we find that the Hubbard $U$ corrections mainly affect the electronic
structure through a moderate modification of the Fermi surface, while
leaving the primary Ni-$d_{x^2-y^2}$ band largely intact. This results
in only a limited increase in the available scattering phase space,
suggesting that the electron–phonon coupling is unlikely to be
strongly enhanced by correlation effects through the electronic
channel alone, as discussed in the following sections.


\begin{figure}[t]  
\centering  
\includegraphics[width=0.7\textwidth]{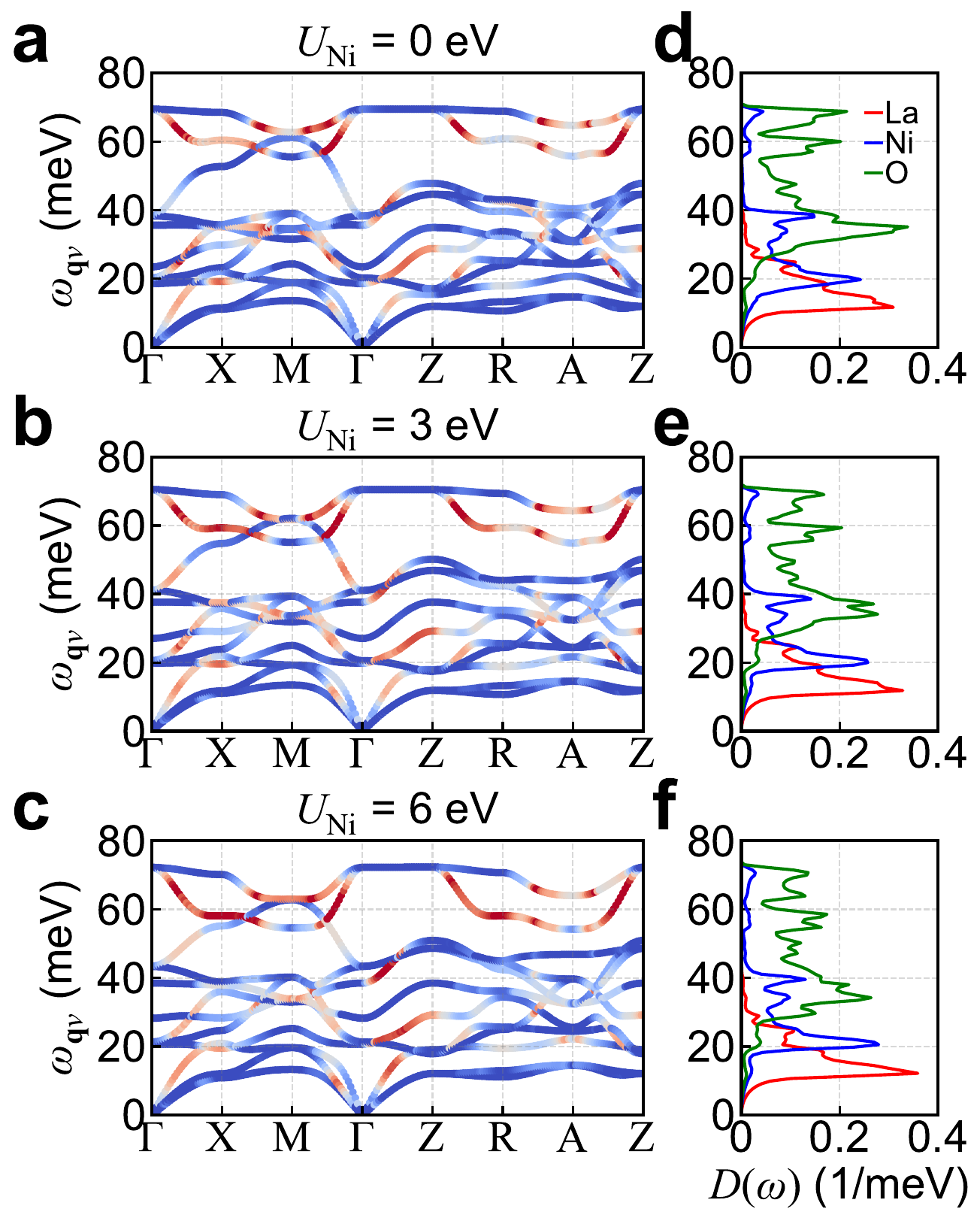}
\caption{Phonon properties of 20\% hole-doped LaNiO$_2$. (a-c) Phonon
  spectrum calculated at $U_{\rm{Ni}}$ = 0, 3, 6 eV, respectively.
  The color scale represents the magnitude of the mode-resolved
  electron–phonon matrix element $g_{\textbf{q}\nu}$, averaged over
  the Fermi surface for each phonon mode $\nu$ and wavevector
  \textbf{q}; red indicates the largest values and blue the smallest.
  (d-f) Phonon density of states calculated at $U_{\rm{Ni}}$ = 0, 3, 6
  eV, respectively. The red, blue and green lines represent the
  contributions from La, Ni and O atoms, respectively.}
\label{fig:fig3}
\end{figure}

Next we study the evolution of the phonon properties of 20\%
hole-doped LaNiO$_2$, calculated at a few representative Hubbard
$U_{\rm{Ni}}$. Figure~\ref{fig:fig3}(a-c) show the phonon spectra of
20\% hole-doped LaNiO$_2$, calculated at different $U_{\rm{Ni}}$
values. Fig.~\ref{fig:fig3}(d-f) show the corresponding atom-resolved
phonon DOS. We use the tetragonal structure with space group $P4/mmm$
to perform the phonon calculations~\cite{PhysRevB.105.115134}. We find
that they are all dynamically stable in the $P4/mmm$ structure,
regardless of whether Hubbard $U_{\rm{Ni}}$ corrections are
applied. Furthermore, Hubbard $U_{\rm{Ni}}$ corrections have almost
negligible effects on the phonon band structure and the phonon DOS of
20\% hole-doped LaNiO$_2$, as shown in Fig.~\ref{fig:fig3}. This
indicates that the lattice dynamics of LaNiO$_2$ are largely insensitive
to Hubbard $U$ corrections, implying that correlation effects do not
significantly modify electron–phonon coupling through the phonon
channel.

Equipped with the electronic and phonon band structures, we now study
the electron-phonon properties of 20\% hole-doped LaNiO$_2$ and
their dependence on Hubbard $U_{\rm{Ni}}$. In Figure~\ref{fig:fig3}(a-c),
we show the mode-resolved electron-phonon matrix $g_{\textbf{q}\nu}$
by color on the phonon spectrum. The mode-resolved
electron-phonon matrix $g_{\textbf{q}\nu}$ is defined as follows:
\begin{equation}
\label{eq1}  
 g_{\textbf{q}\nu}=\frac{1}{N_bN(\varepsilon_F)}\sqrt{\sum_{m,n}\int_{\rm{BZ}}\frac{d\mathbf{k}}{V_{\rm{BZ}}}|g_{mn,\nu}(\mathbf{k},\mathbf{q})|^2 \delta(\varepsilon_{m\mathbf{k}}-\varepsilon_F) \delta(\varepsilon_{n\mathbf{k}+\mathbf{q}}-\varepsilon_F)}
\end{equation}
where $N_b$ is the number of bands in the Wannier fitting for
electron-phonon calculations. $\varepsilon_F$ is the Fermi level and
$N(\varepsilon_F)$ is DOS at the Fermi
level. $V_{\rm{BZ}}$ is the volume of the first electronic Brillouin zone
(BZ). $\varepsilon_{n\textbf{k}}$ is the electronic band energy and
$g_{mn,\nu}(\textbf{k},\textbf{q})$ is the electron-phonon $g$
matrix. $\textbf{q}$ and $\nu$ are the phonon momentum and phonon
index, respectively. The color scale is proportional to the
magnitude of $g_{\textbf{q}\nu}$ with red indicating the largest values
and blue the smallest. Fig.~\ref{fig:fig3}(a-c) show that
all the phonon branches have a sizable $g_{\textbf{q}\nu}$ and no
single phonon mode has a dominant contribution. This indicates that a
full sampling of the entire phonon BZ is necessary when one calculates the
electron-phonon interactions of 20\% hole-doped LaNiO$_2$. 




\begin{figure}[t]  
\centering  
\includegraphics[width=0.98\textwidth]{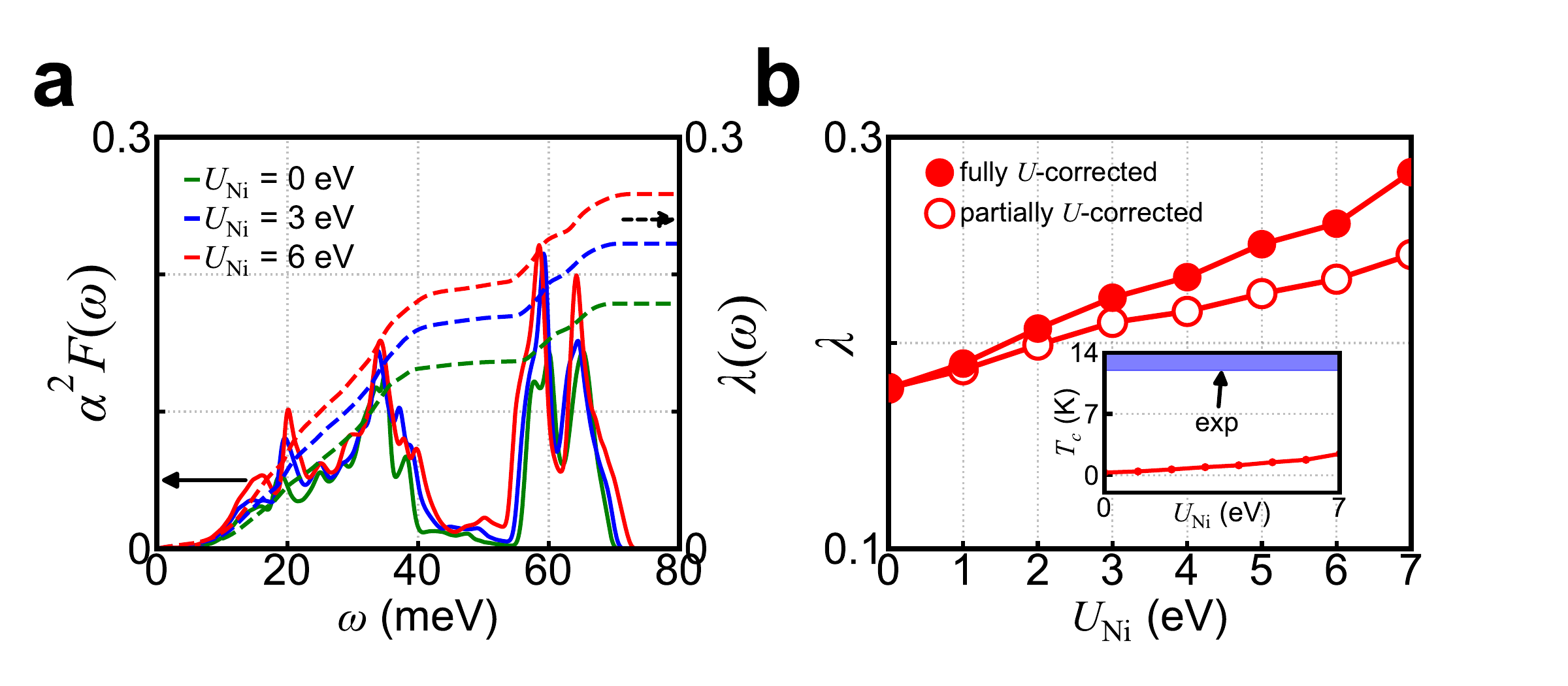}  
\caption{Electron-phonon properties of 20\% hole-doped LaNiO$_2$. (a)
  Electron-phonon spectral function $\alpha^2F(\omega)$ and
  accumulative electron-phonon coupling $\lambda$($\omega$) calculated
  at $U_{\rm{Ni}}$ = 0, 3, 6 eV. The solid curves represent
  $\alpha^2F(\omega)$ and the dashed curves correspond to
  $\lambda$($\omega$). (b) Total electron-phonon coupling $\lambda$ as
  a function of $U_{\rm{Ni}}$. The solid and open symbols represent
  the total electron-phonon coupling calculated by the ``fully
  $U$-corrected'' and ``partially $U$-corrected'' approaches,
  respectively. The inset shows the superconducting temperature $T_c$
  as a function of $U_{\rm{Ni}}$. The purple area highlights the range
  of experimentally observed superconducting transition temperature.}
\label{fig:fig4}
\end{figure}

In Figure~\ref{fig:fig4}, we show more details of the electron-phonon
properties of 20\% hole-doped LaNiO$_2$. In Fig.~\ref{fig:fig4}(a),
the solid curves show the evolution of the electron-phonon spectrum function
$\alpha^2F(\omega)$ of 20\% hole-doped LaNiO$_2$ as a function of
$U_{\rm{Ni}}$, and the dashed curves are the corresponding
accumulative electron-phonon coupling $\lambda(\omega)$.
$\alpha^2F(\omega)$ is defined as follows~\cite{PhysRevB.87.024505}:
\begin{equation}
\label{eq2}  
    \alpha^2F(\omega)=\frac{1}{2N(\varepsilon_F)}\sum_{mn,\nu}\int_{\rm{BZ}}\frac{d\mathbf{k}}{V_{\rm{BZ}}}\frac{d\mathbf{q}}{\Omega_{\rm{BZ}}}|g_{mn,\nu}(\mathbf{k},\mathbf{q})|^2 \delta(\varepsilon_{m\mathbf{k}}-\varepsilon_F) \delta(\varepsilon_{n\mathbf{k}+\mathbf{q}}-\varepsilon_F)\delta(\omega-\omega_{\mathbf{q}\nu})
\end{equation}
where $V_{\rm{BZ}}$ and $\Omega_{\rm{BZ}}$ are the volume of the first electronic
and phonon BZ, respectively. $\varepsilon_{n\textbf{k}}$ and
$\omega_{\textbf{q}\nu}$ are the electronic and phonon band energies. Based on
$\alpha^2F(\omega)$, the accumulative electron-phonon coupling
$\lambda(\omega)$ is defined as:
\begin{equation}
\label{eq3}\lambda(\omega)=2\int^{\omega}_0\frac{\alpha^2F(\omega')}{\omega'}d\omega'
\end{equation}
When $\omega$ tends to infinity, $\lambda(\omega)$ yields the total electron-phonon coupling $\lambda$. In Fig.~\ref{fig:fig4}(a), we find that there are four peaks in $\alpha^2F(\omega)$. Among these four peaks, Hubbard $U_{\rm{Ni}}$ corrections slightly enhance the two high-frequency peaks that are located around 59 meV and 64 meV. Those two peaks are mainly associated with oxygen vibrations. In Fig.~\ref{fig:fig4}(b), the solid symbols show the evolution of the total electron-phonon coupling $\lambda$ as a function of Hubbard $U_{\rm{Ni}}$. We find that Hubbard $U_{\rm{Ni}}$ corrections increase $\lambda$ from 0.18 at $U_{\rm{Ni}}=0$ to 0.28 at $U_{\rm{Ni}} = 7$ eV. However, the magnitude of $\lambda$ still remains small. Using the McMillan-Allen-Dynes formula and setting the Coulomb pseudopotential to zero (the most favorable setting for superconducting transition temperature $T_c$), we estimate that $T_c$ does not exceed 3 K for any reasonable value of $U_{\rm{Ni}}$, as shown in the inset of Fig.~\ref{fig:fig4}(b). By contrast, the experimental superconducting transition temperature $T_c$ of 20\% hole-doped LaNiO$_2$ is around 12-14 K, highlighted by the purple shade in the inset of Fig.~\ref{fig:fig4}(b). This indicates that even fully taking into account the Hubbard $U$ corrections, the electron-phonon interactions alone are not sufficient to explain the superconductivity in infinite-layer nickelates. 

In passing, we comment on an important technical subtlety: we apply Hubbard $U$ corrections not only to electronic structure, but also to phonon spectrum and electron-phonon $g$ matrix. We term this approach as ``fully $U$-corrected''. This is similar to Ref.~\cite{stevein_twogap}, in which full GW corrections are applied throughout the electron-phonon calculations. In addition, there is another approximate approach for calculating electron-phonon interaction by only considering correlation corrections (Hubbard $U$ or GW) in electronic structure, while using standard DFT-calculated phonon spectrum and electron-phonon $g$ matrix~\cite{Preempted}. In this treatment, if correlation effects are modelled as ``Hubbard $U$ corrections'', we term it as ``partially $U$-corrected''. In Fig.~\ref{fig:fig4}(b), we compare the total electron-phonon coupling $\lambda$ calculated by using the ``partially $U$-corrected'' approach (open symbols) to the ``fully $U$-corrected'' approach (solid symbols). We find that qualitatively the two approaches yield the same trend, but quantitatively the ``partially $U$-corrected'' approach underestimates the total electron-phonon coupling $\lambda$. For the next material RuO$_2$, the discrepancy between the two approaches becomes more pronounced. Therefore for a self-consistent and quantitatively accurate calculation of electron-phonon interactions for real correlated materials, correlation corrections (regardless of Hubbard $U$ or GW) should be considered not only in electronic structure, but also in phonon spectrum and electron-phonon $g$ matrix.  


\begin{figure}[t]  
\centering  
\includegraphics[width=0.95\textwidth]{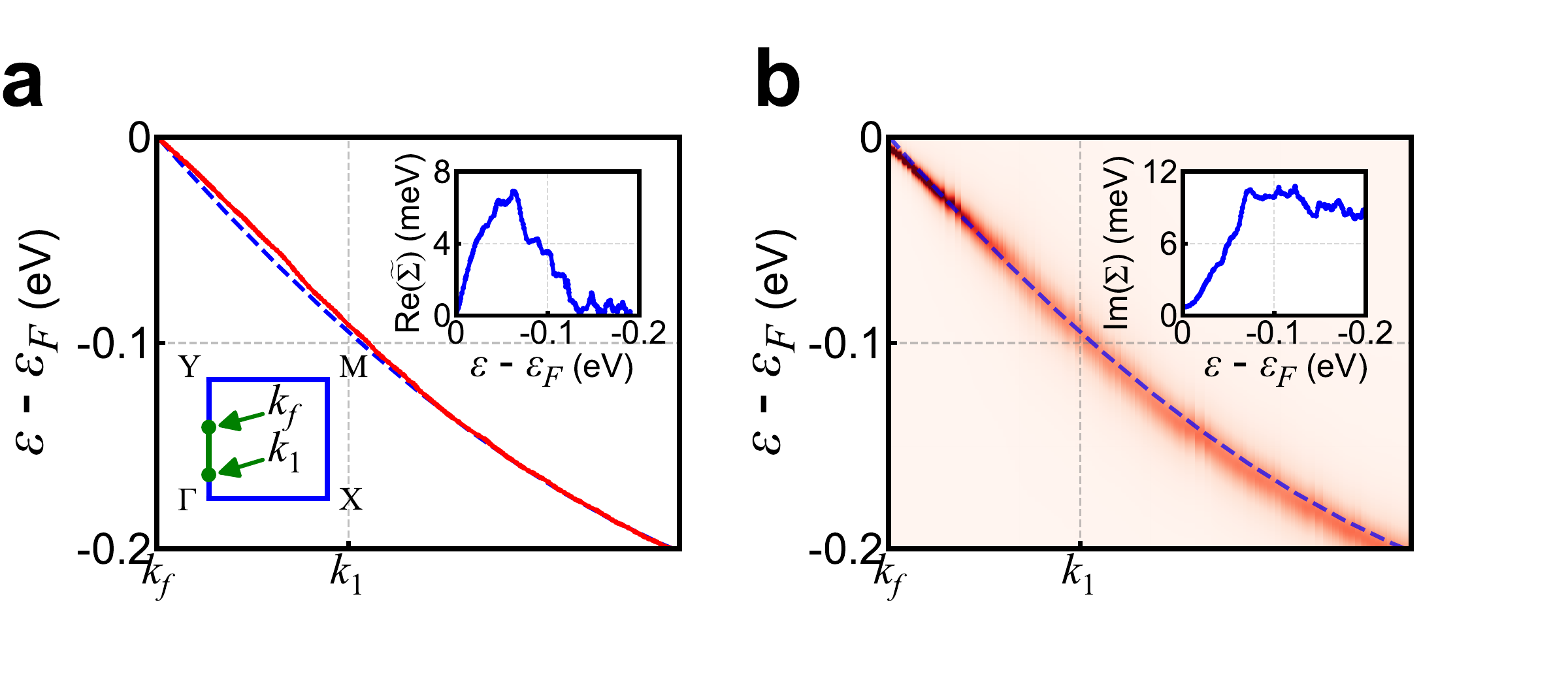}  
\caption{Electronic self-energy of 20\% hole-doped LaNiO$_2$ that arises
  from the electron-phonon interactions, calculated at $U_{\rm{Ni}}$ =
  6 eV.  (a) Comparison between quasiparticle energy (red solid line)
  and bare energy (blue dashed line). The left inset shows the
  high-symmetry $\mathbf{k}$ path. $k_f$ is the Fermi wave vector and $k_1$
  is the wave vector corresponding to $\varepsilon =\varepsilon_F
  -0.1$ eV. The right inset shows the real part of the electronic
  self-energy between $\varepsilon_F$ and $\varepsilon_F-0.2$ eV. (b)
  Spectral function $A_{n\textbf{k}}(\omega,T)$ associated with the
  electronic self-energy. The blue dashed line represents the bare
  energy dispersion $\varepsilon_{n\textbf{k}}$. The inset shows the
  imaginary part of the electronic self-energy between $\varepsilon_F$
  and $\varepsilon_F-0.2$ eV. The temperature $T$ is set to 20 K.}
\label{fig:fig5}
\end{figure}


Next we study another important observable: the electronic self-energy of 20\% hole-doped LaNiO$_2$ that arises from electron-phonon interactions, calculated at a representative $U_{\rm{Ni}}$ = 6 eV. The electronic self-energy $\Sigma_{n\mathbf{k}}(\omega,T)$ is defined as~\cite{RevModPhys.89.015003}:
\begin{multline}
  \label{eq4}
  \Sigma_{n\mathbf{k}}(\omega,T)=\sum_{m\nu}\int_{\rm{BZ}}\frac{d\mathbf{q}}{\Omega_{\rm{BZ}}}|g_{mn,\nu}(\mathbf{k},\mathbf{q})|^2   \\
  \times  \left[ \frac{n_{\mathbf{q}\nu}(T)+f_{m\mathbf{k}+\mathbf{q}}(T)}{\omega-(\varepsilon_{m\mathbf{k}+\mathbf{q}}-\varepsilon_F)+\omega_{\mathbf{q}\nu}+i\delta}  +  \frac{n_{\mathbf{q}\nu}(T)+1-f_{m\mathbf{k}+\mathbf{q}}(T)}{\omega-(\varepsilon_{m\mathbf{k}+\mathbf{q}}-\varepsilon_F)-\omega_{\mathbf{q}\nu}+i\delta} \right]     
\end{multline}
where
$n_{\mathbf{q}\nu}(T)=\left[\textrm{exp}\left(\frac{\omega_{\mathbf{q}\nu}}{k_BT}\right)-1\right]^{-1}$
is the Bose-Einstein distribution and $f_{n\mathbf{k}}(T) =
\left[\textrm{exp}\left(\frac{\varepsilon_{n\textbf{k}}-\mu}{k_BT}\right)+1\right]^{-1}$
is the Fermi-Dirac distribution. $k_B$ is the Boltzmann constant;
$\mu$ is the chemical potential; $T$ is the temperature.  $\delta$ is
a small positive number which guarantees the correct analytical
structure of the self-energy, and also avoids numerical
instabilities. Throughout the calculations, we set $\delta $ = 3 meV
and the temperature $T$ to 20 K. Since 20\% hole-doped LaNiO$_2$ is
metallic, one needs to contain a Debye-Waller term Re$
\Sigma_{n\mathbf{k}}(\omega=\varepsilon_{F},T)$ in the electronic
self-energy~\cite{RevModPhys.89.015003}. That is:
\begin{equation}
\label{eq5}  
   \widetilde{\Sigma}_{n\mathbf{k}}(\omega,T)=\Sigma_{n\mathbf{k}}(\omega,T)-\text{Re}\Sigma_{n\mathbf{k}}(\omega=\varepsilon_{F},T)
\end{equation}
The Debye-Waller term is always real and static, which guarantees that
the real part of the corrected electronic self-energy vanishes at the
Fermi level,
i.e. $\textrm{Re}\widetilde{\Sigma}_{n\mathbf{k}}(\varepsilon_F,T)=0$~\cite{PhysRevLett.126.146401}.
We note that the Debye-Waller term does not affect the imaginary part
and thus for simplicity we use the notation
$\textrm{Im}\Sigma_{n\textbf{k}}(\omega, T)$ instead of
$\textrm{Im}\widetilde{\Sigma}_{n\textbf{k}}(\omega, T)$.

In Figure~\ref{fig:fig5}, we study the physical manifestation of the
electronic self-energy $\widetilde{\Sigma}_{n\mathbf{k}}(\omega,T)$. In
Fig.~\ref{fig:fig5}(a), the blue dashed line is the bare energy
dispersion $\varepsilon_{n\textbf{k}}$ around the Fermi level along
$\Gamma$ to X. By comparison, the red solid line represents the
quasiparticle energy dispersion $E_{n\textbf{k}}$ dressed by the
electron-phonon interactions:
\begin{equation}
\label{eq6}  
E_{n\textbf{k}}=\varepsilon_{n\textbf{k}} + \text{Re}\widetilde{\Sigma}_{n\mathbf{k}}(\varepsilon_{n\textbf{k}},T)
\end{equation}
The inset shows the real part of the corrected self-energy Re$ \widetilde{\Sigma}_{n\mathbf{k}}(\omega,T)$ in the energy range from $\varepsilon_F$ to $\varepsilon_F-0.2$ eV. We find that in 20\% hole-doped LaNiO$_2$, the electron-phonon interactions renormalize the bare energy dispersion and result in a ``kink'' feature around 64 meV below the Fermi level. This ``kink'' feature is reminiscent of cuprates, which also exhibit similar spectroscopic signature around 70 meV below the Fermi level~\cite{PhysRevLett.126.146401}. However, this ``kink'' in infinite-layer nickelates is substantially smaller than that in cuprates, indicating much weaker electron-phonon interactions in infinite-layer nickelates. In Fig.~\ref{fig:fig5}(b), we show the color map of the interacting spectral function $A_{n\mathbf{k}}(\omega, T)$, which is defined as~\cite{EPW}:
\begin{equation}
  \label{eq7}
  A_{n\mathbf{k}}(\omega, T)
  = \frac{1}{\pi}
  \frac{|\text{Im}\Sigma_{n\mathbf{k}}(\omega, T)|}
  {|\omega - (\varepsilon_{n\mathbf{k}} - \varepsilon_F) - \text{Re}\widetilde{\Sigma}_{n\mathbf{k}}(\omega, T)|^2 + |\text{Im}\Sigma_{n\mathbf{k}}(\omega, T)|^2 }
\end{equation}
Because scattering rate and inverse life time of electrons are proportional to the imaginary part of electronic self-energy, the spectral function shows that the electron-phonon interactions make the quasiparticle energy dispersion $E_{n\textbf{k}}$ blurred to some extent. The inset of Fig.~\ref{fig:fig5}(b) shows the imaginary part of the electronic self-energy Im$\Sigma_{n\mathbf{k}}(\omega, T)$ in the energy range from $\varepsilon_F$ to $\varepsilon_F-0.2$ eV. Similar to the real part, the imaginary part of the electronic self-energy also remains small, on the order of 10 meV.


  
\subsection{Ruthenium dioxide}


In this section, we study ruthenium dioxide (RuO$_2$), in both its bulk and strained forms. As mentioned in the Introduction, RuO$_2$ thin films strained on TiO$_2$ substrates show superconductivity below 1.5 K. For clarity, we present the results of TiO$_2$-strained RuO$_2$ in the main text and the results of bulk RuO$_2$ in the Supplementary Note 3 that includes Fig. S7-S11. For RuO$_2$, the onsite Hubbard $U$ interaction is applied to Ru-$d$ orbitals.

Both bulk RuO$_2$ and TiO$_2$ crystallize in the orthorhombic
structure of space group $P4_2/mnm$ (No. 136). When strained to TiO$_2$,
RuO$_2$ has an anisotropic lattice mismatch of +2.3\% along the
[1$\overline{1}$0] direction and -3.1\% in the [001] direction. This
leads to a new crystal structure of space group $Pnnm$ (No. 58) for
strained RuO$_2$.


\begin{figure}[ht]
\centering \includegraphics[width=0.95\textwidth]{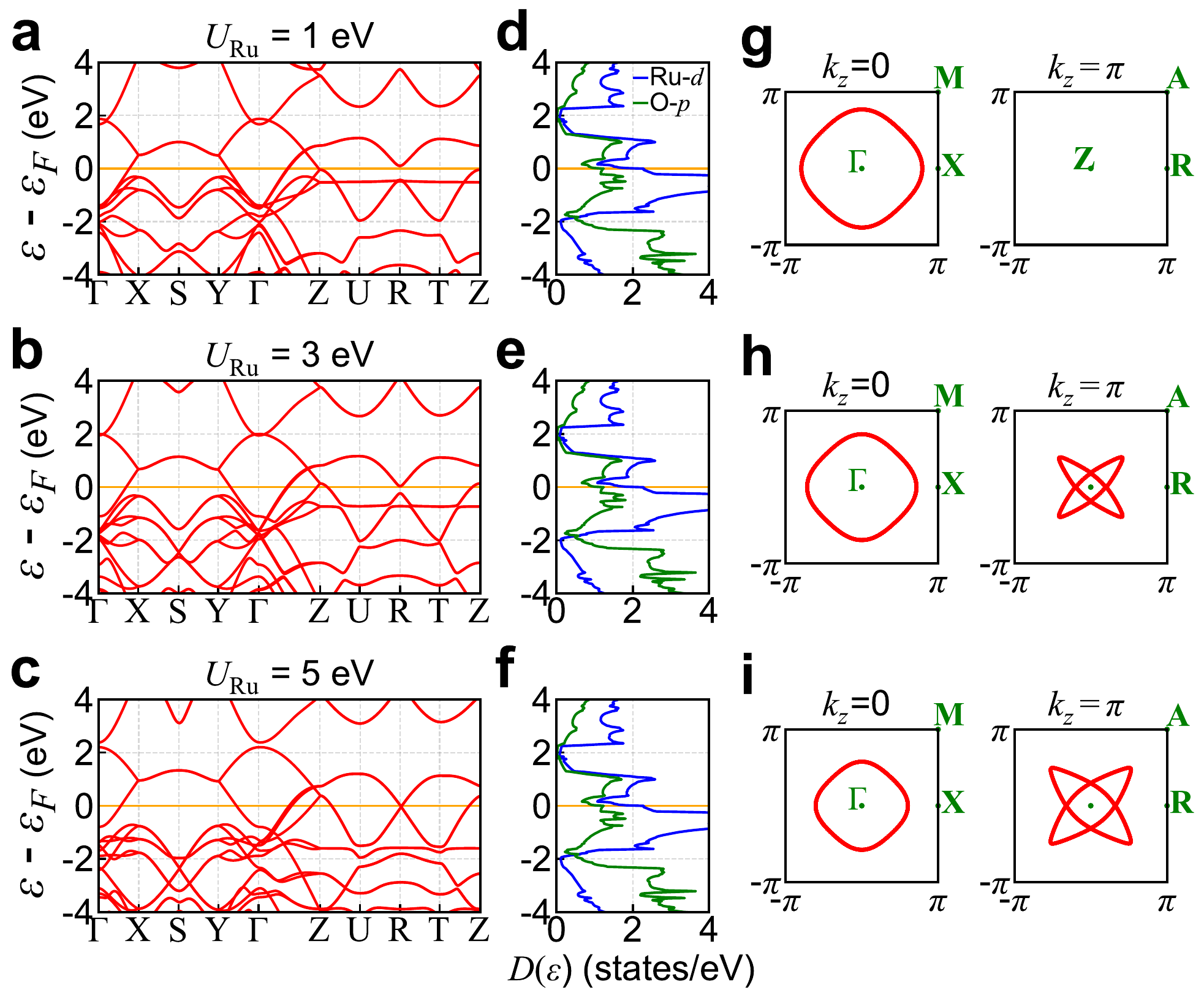}
\caption{Electronic properties of strained RuO$_2$: (a-c) Band structure
  calculated at $U_{\rm{Ru}}$ = 1, 3, 5 eV, respectively. (d-f)
  Density of states calculated at $U_{\rm{Ru}}$ = 1, 3, 5 eV,
  respectively.  The blue and green lines denote the projections onto
  Ru-$d$ and O-$p$ orbitals, respectively.  (g-i) Fermi surface
  calculated at $U_{\rm{Ru}}$ = 1, 3, 5 eV, respectively. Left
  sub-panel is for $k_z$ = 0 plane and right sub-panel is for $k_z$ =
  $\pi$ plane, respectively.}
\label{fig:fig6}
\end{figure}

\begin{figure}[ht]  
\centering  
\includegraphics[width=0.7\textwidth]{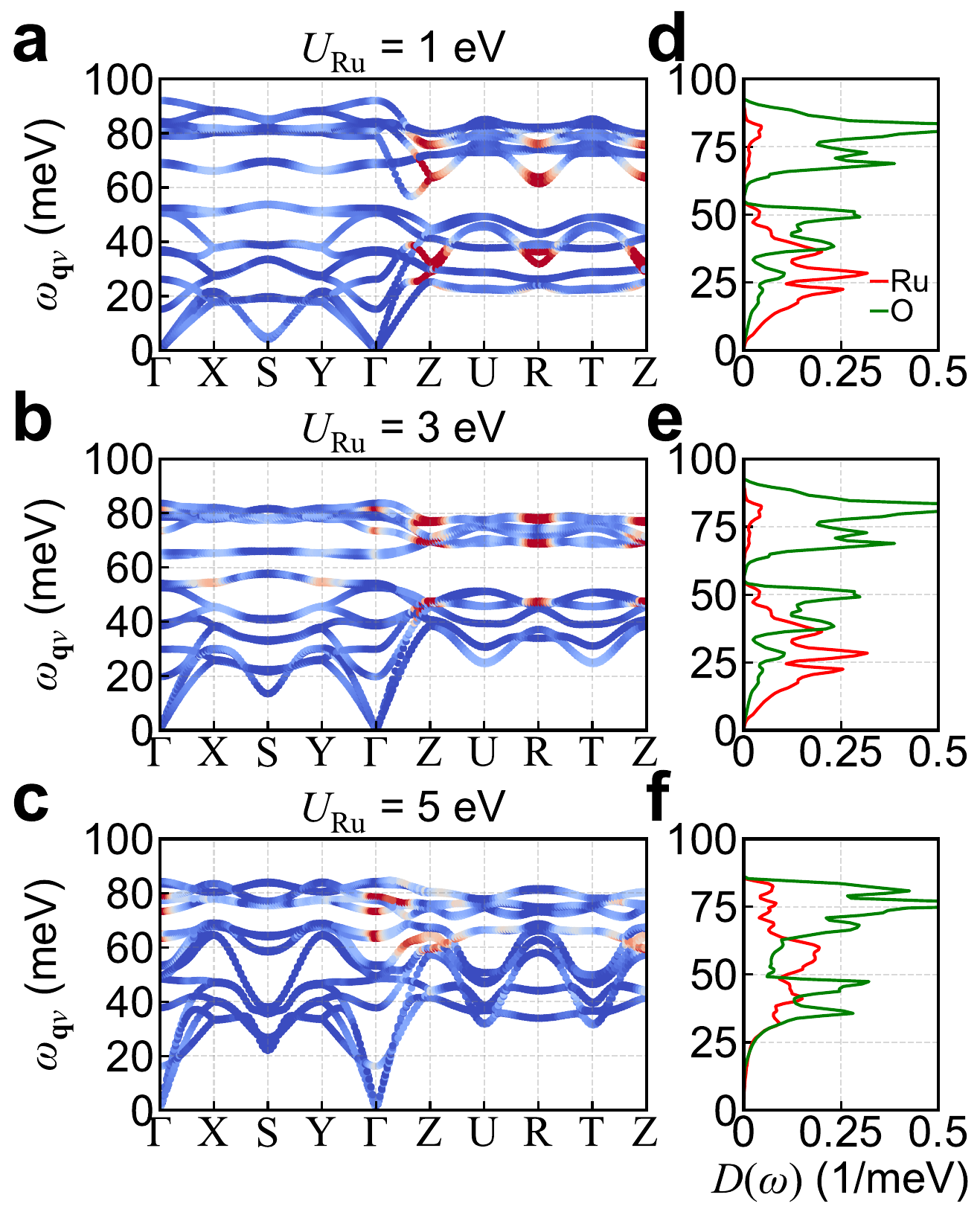}
\caption{Phonon properties of strained RuO$_2$. (a-c) Phonon spectrum
  calculated at $U_{\rm{Ru}}$ = 1, 3, 5 eV, respectively.  The color
  scale represents the magnitude of the mode-resolved electron–phonon
  matrix element $g_{\textbf{q}\nu}$, averaged over the Fermi surface
  for each phonon mode $\nu$ and wavevector \textbf{q}; red indicates
  the largest values and blue the smallest.  (d-f) Phonon density of
  states calculated at $U_{\rm{Ru}}$ = 1, 3, 5 eV, respectively. The
  red and green lines represent the contributions from Ru and O atoms,
  respectively.}
\label{fig:fig7}
\end{figure}

Figure~\ref{fig:fig6}(a-c) and (d-f) show the evolution of the
electronic band structure and DOS of TiO$_2$-strained RuO$_2$,
calculated at a few representative Hubbard
$U_{\rm{Ru}}$. We find that for a physically reasonable range of
$U_{\rm{Ru}}$, the states around the Fermi level are dominantly
derived from Ru-$d$ and O-$p$ orbitals. Compared to infinite-layer
nickelates, the metal-oxygen hybridization between Ru-$d$ and O-$p$
orbitals is substantially stronger in strained
RuO$_2$. Fig.~\ref{fig:fig6}(g-i) show the corresponding Fermi surface
at $k_z=0$ and $k_z=\pi$ planes. We find that the Fermi surface of
strained RuO$_2$ is also strongly
three-dimensional. With an increasing $U_{\rm{Ru}}$, the large
electron pocket around $\Gamma$ slightly shrinks, while a hole pocket
around Z point emerges (as $U_{\rm{Ru}} > 1$ eV) and expands.


Next we present the evolution of the phonon spectrum and the
atom-resolved phonon DOS of strained RuO$_2$, calculated at a few
representative Hubbard $U_{\rm{Ru}}$ in Figure~\ref{fig:fig7}. As
$U_{\rm{Ru}}$ increases from 1 eV to 5 eV, the phonon spectrum is free
from imaginary phonon modes, indicating that the $Pnnm$ orthorhombic
structure is dynamically stable. However, we need to mention that in
standard DFT calculations (i.e. at $U_{\rm{Ru}} = 0$), there are
imaginary phonon modes at S and Z points (see Supplementary Note 4 and
Fig. S12 therein). Considering that the $Pnnm$ structure is
experimentally observed and stable~\cite{PhysRevLett.133.176401}, this
indicates that correlation effects such as Hubbard $U_{\rm{Ru}}$
corrections are essential in describing the phonon properties of
RuO$_2$. As $U_{\rm{Ru}}$ is increased, there is a clear tendency of
phonon hardening. In particular, the phonon frequency of the lowest
mode at S (Z) point increases from 4 (25) meV at $U_{\rm{Ru}} = 1$ eV
to 22 (41) meV at $U_{\rm{Ru}} = 5$ eV. This correlation-driven phonon
hardening effect will play an important role in the electron-phonon
properties of RuO$_2$. In passing, we would like to
  mention that in finite-temperature phonon calculations, the
  imaginary phonon modes of strained RuO$_2$ may also be eliminated by
  anharmonic effects. However, the focus of the present work is
  superconductivity in strained RuO$_2$, which emerges at the low
  transition temperature of $T_c$ = 1.5 K. The experimentally observed
  orthorhombic structure remains dynamically stable down to low temperatures.
  Therefore we employ standard zero temperature phonon calculations within
  harmonic approximation. More details on this point are found in the
  Supplementary Note 5 that includes Fig. S13.

Equipped with the electronic and phonon band structures, we now study
the electron-phonon properties of strained RuO$_2$ and their
dependence on Hubbard $U_{\rm{Ru}}$. In Fig.~\ref{fig:fig7}(a-c), we
show the mode-resolved electron-phonon matrix $g_{\textbf{q}\nu}$ by
color on the phonon spectrum. The mode-resolved electron-phonon matrix
$g_{\textbf{q}\nu}$ is defined in Eq.~(\ref{eq1}). The color scale is
proportional to the magnitude of $g_{\textbf{q}\nu}$ with red
indicating the largest values and blue the smallest.  We find that
similar to infinite-layer nickelates, all the phonon branches in
strained RuO$_2$ have a sizable $g_{\textbf{q}\nu}$ and no single
phonon mode has a dominant contribution.

\begin{figure}[t]  
\centering  
\includegraphics[width=0.98\textwidth]{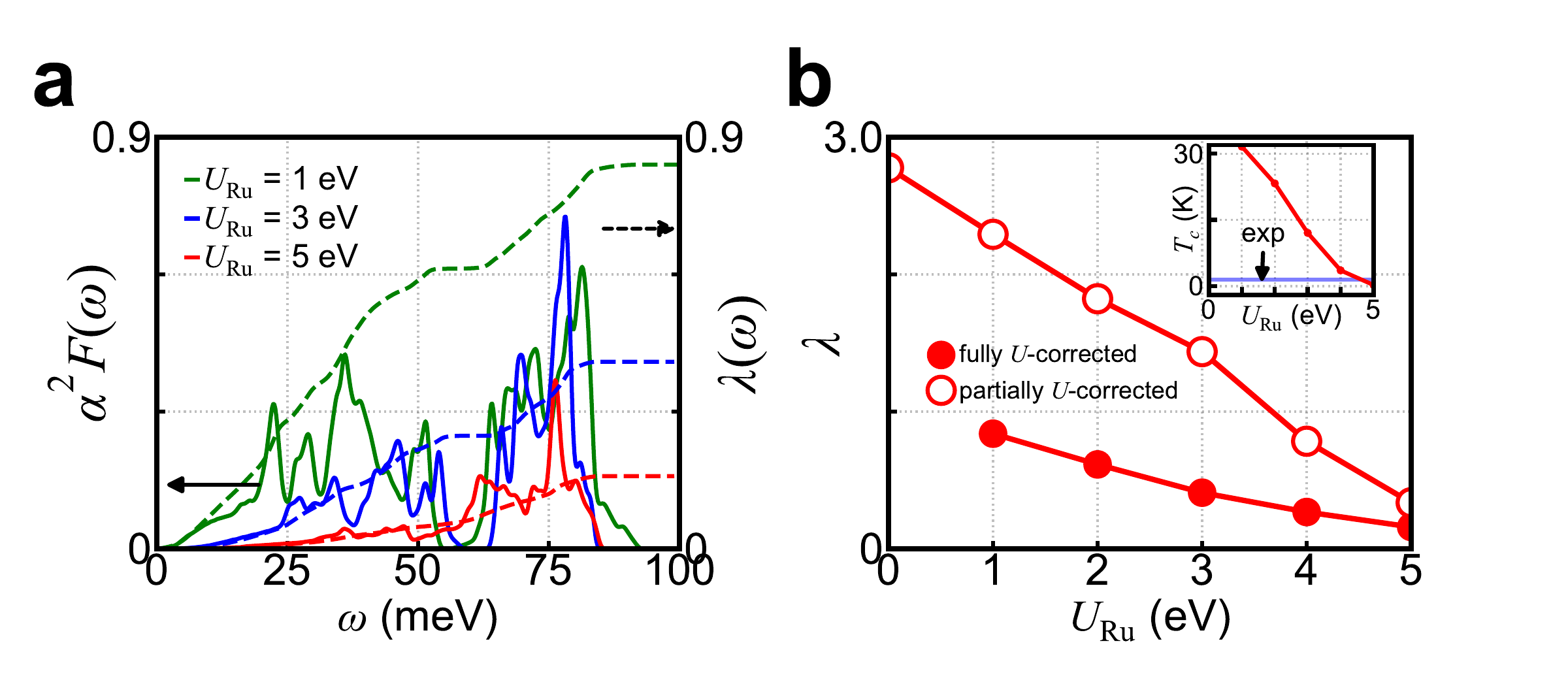}  
\caption{Electron-phonon properties of strained RuO$_2$. (a)
  Electron-phonon spectral function $\alpha^2F(\omega)$ and
  accumulative electron-phonon coupling $\lambda$($\omega$) calculated
  at $U_{\rm{Ru}}$ = 1, 3, 5 eV. The solid curves represent
  $\alpha^2F(\omega)$ and the dashed curves correspond to
  $\lambda$($\omega$). (b) Total electron-phonon coupling $\lambda$ as
  a function of $U_{\rm{Ru}}$. The solid and open symbols represent
  the total electron-phonon coupling calculated by the ``fully
  $U$-corrected'' and ``partially $U$-corrected'' approaches,
  respectively. The inset shows the superconducting temperature $T_c$
  as a function of $U_{\rm{Ru}}$. The purple line highlights the
  experimentally observed superconducting transition temperature. We
  note that in the ``fully $U$-corrected'' calculations, we start from
  $U_{\rm{Ru}} \ge 1$ eV to ensure that the experimentally observed
  $Pnnm$ structure is dynamically stable.}
\label{fig:fig8}
\end{figure}

In Figure~\ref{fig:fig8}, we show more details of the electron-phonon
properties of strained RuO$_2$. In Fig.~\ref{fig:fig8}(a), the solid
curves show the evolution of the electron-phonon spectrum function
$\alpha^2F(\omega)$ of RuO$_2$, calculated at a few representative
Hubbard $U_{\rm{Ru}}$, and the dashed curves are the corresponding
accumulative electron-phonon coupling
$\lambda(\omega)$. $\alpha^2F(\omega)$ and $\lambda(\omega)$ are
defined in Eq.~(\ref{eq2}) and Eq.~(\ref{eq3}), respectively. In
contrast to infinite-layer nickelates, the correlation effects
modelled by Hubbard $U$ corrections substantially reduce the
electron-phonon spectrum function $\alpha^2F(\omega)$ of
RuO$_2$. Consequently, the total electron-phonon coupling $\lambda$
decreases with $U_{\rm{Ru}}$ from 0.8 at $U_{\rm{Ru}} = 1$ eV to 0.2
at $U_{\rm{Ru}} = 5$ eV, as shown by the solid symbols in
Fig.~\ref{fig:fig8}(b). In the inset of Fig.~\ref{fig:fig8}(b), we
estimate the superconducting transition temperature $T_c$ by using the
McMillan-Allen-Dynes formula and setting the Coulomb pseudopotential
to zero. As we mentioned previously, the experimentally observed
$Pnnm$ crystal structure is dynamically unstable in standard DFT
calculations (i.e $U_{\rm{Ru}} = 0$). At $U_{\rm{Ni}} = 1$ eV, the
orthorhombic $Pnnm$ structure becomes dynamically stable but its total
electron-phonon coupling is as large as 0.8, which leads to a $T_c$ of
about 30 K. The overestimation of $T_c$ originates from soft phonon
modes, which enhance the electron–phonon coupling. As $U_{\rm{Ru}}$
increases to 4 eV, the estimates $T_c$ is reduced to about a few
Kelvin, which approaches the experimental value of 1.5 K (highlighted
by the purple line). Combining the phonon properties
Fig.~\ref{fig:fig7} and the electron–phonon coupling results
Fig.~\ref{fig:fig8}, we find that the suppression of electron–phonon
coupling in RuO$_2$ is primarily driven by correlation-induced phonon
hardening. In particular, the upward shift of low-frequency phonon
modes significantly reduces their contribution to $\lambda$,
indicating that correlation effects act predominantly through the
phonon channel in this system.


We also briefly comment on the difference between the ``fully
$U$-corrected'' and ``partially $U$-corrected'' approaches. Different
from infinite-layer nickelates, the ``partially $U$-corrected''
approach substantially overestimates the total electron-phonon
coupling $\lambda$ of strained RuO$_2$. This is because in the
``partially $U$-corrected'' calculations, there are imaginary phonon
modes for the observed $Pnnm$ structure from standard DFT calculations.
Hence, the total electron-phonon coupling is substantially and
artificially enhanced because many phonon modes have unphysically low
frequencies. This comparison demonstrates that neglecting correlation
effects in phonon properties can lead to qualitatively incorrect
predictions of electron–phonon coupling. A fully self-consistent
treatment of correlation in both electronic and lattice degrees of
freedom is therefore essential for reliable modeling of
electron-phonon properties of correlated materials.


\begin{figure}[t]  
\centering  
\includegraphics[width=0.95\textwidth]{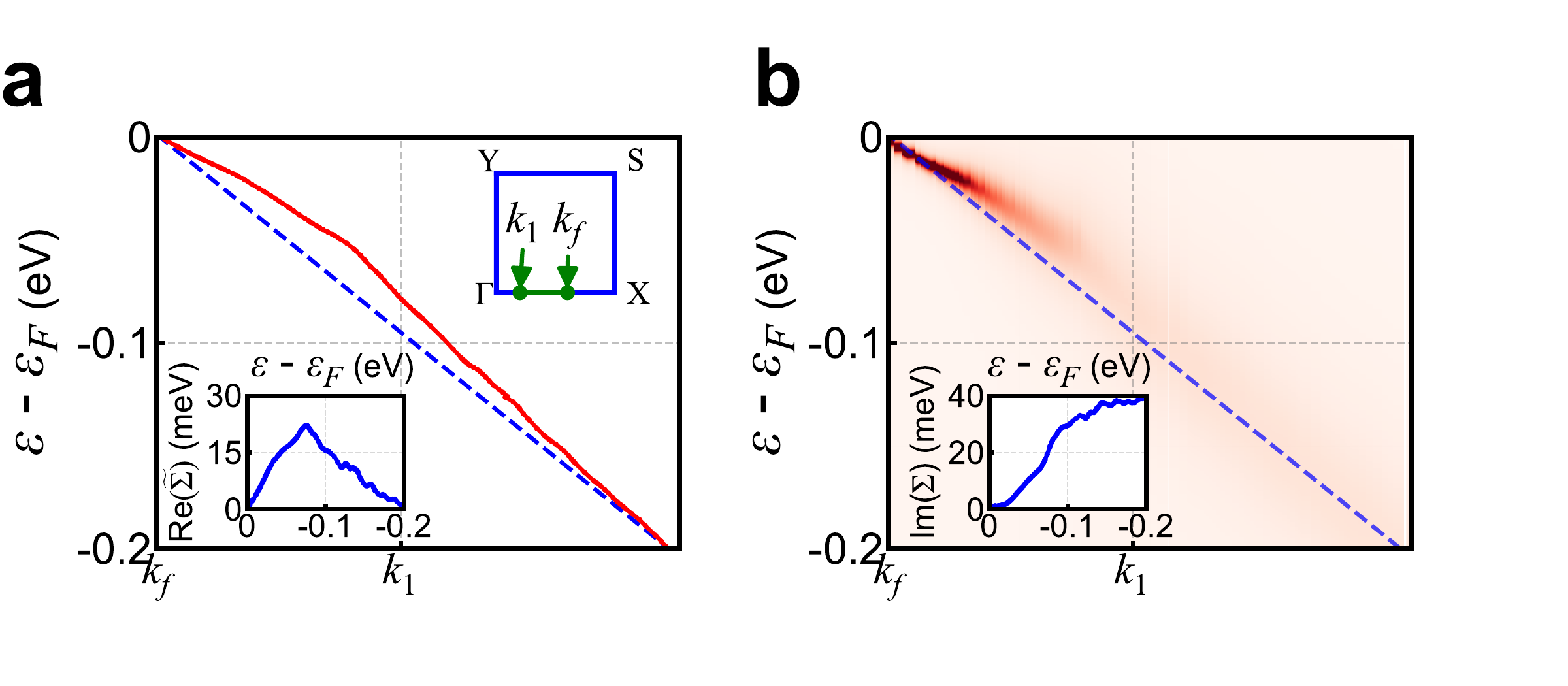}  
\caption{Electronic self-energy of strained RuO$_2$ that arises
  from the electron-phonon interactions, calculated at $U_{\rm{Ru}}$ =
  3 eV. (a) Comparison between quasiparticle energy (red solid line)
  and bare energy (blue dashed line). The right inset shows the
  high-symmetry $\mathbf{k}$ path. $k_f$ is the Fermi wave vector and $k_1$
  is the wave vector corresponding to $\varepsilon =\varepsilon_F
  -0.1$ eV. The left inset shows the real part of the electronic
  self-energy between $\varepsilon_F$ and $\varepsilon_F-0.2$ eV. (b)
  Spectral function $A_{n\textbf{k}}(\omega,T)$ associated with the
  electronic self-energy. The blue dashed line represents the bare
  energy dispersion $\varepsilon_{n\textbf{k}}$. The inset shows the
  imaginary part of the electronic self-energy between $\varepsilon_F$
  and $\varepsilon_F-0.2$ eV. The temperature $T$ is set to 20 K.}
\label{fig:fig9}
\end{figure}

Next we study the electronic self-energy of strained RuO$_2$ that arises from electron-phonon interactions calculated at a representative $U_{\rm{Ru}} = 3$ eV. Figure~\ref{fig:fig9}(a) shows the quasiparticle energy $E_{n\textbf{k}}$ defined in Eq.~(\ref{eq6}) by the red solid line. The DFT-calculated bare energy is shown by the blue dashed line in comparison. The high-symmetry $\textbf{k}$ path is chosen along $\Gamma$ to X. The Debye-Waller term is included so that the quasiparticle energy $E_{n\textbf{k}}$ is identical to the bare energy $\varepsilon_{n\textbf{k}}$ at the Fermi level. Compared to infinite-layer nickelates, the real part of the electronic self-energy of strained RuO$_2$ is substantially larger, as shown in the inset of Fig.~\ref{fig:fig9}(a). Hence the ``kink'' feature becomes more pronounced. Fig.~\ref{fig:fig9}(b) shows the spectral function $A_{n\textbf{k}}(\omega, T)$ defined in Eq.~(\ref{eq7}) and the inset shows the imaginary part of the electronic self-energy of strained RuO$_2$. Similar to the real-part, the imaginary part of the electronic self-energy is also substantially larger than that in infinite-layer nickelates. This leads to strong ``blurring'' effects of quasiparticle dispersion in the spectral function $A_{n\textbf{k}}(\omega, T)$, in particular when the energy is about 50 meV below the Fermi level.

Overall, in contrast to LaNiO$_2$, where correlation effects only weakly modify the electronic structure and have a negligible impact on phonon properties, RuO$_2$ exhibits strong correlation-driven hardening in lattice dynamics, leading to a pronounced suppression of electron–phonon coupling.

\section{Discussion}

Before conclusions, we make some comments.

In this work, we employ a fully self-consistent Hubbard $U$ framework
for electron–phonon interactions in correlated materials, in which
Hubbard $U$ corrections are incorporated not only in the electronic
structure, but also in the phonon spectrum and electron–phonon
coupling matrix elements. This allows us to systematically investigate
how electronic correlation influences electron–phonon coupling in
correlated materials. Our results reveal that correlation effects
modify electron–phonon coupling through two distinct
mechanisms. First, correlation can alter the electronic structure,
including Fermi surface topology and orbital character, thereby
changing the available scattering phase space (electronic
channel). Second, correlation effects can directly renormalize lattice
dynamics, leading to phonon hardening or softening and thus modifying
the phonon contribution to the electron–phonon coupling (phonon
channel). The relative importance of these two mechanisms is strongly
material dependent.

In LaNiO$_2$, our calculations show that Hubbard $U$ corrections lead
to only a marginal enhancement of electron–phonon coupling. This
result stands in contrast to previous GW-based
study~\cite{stevein_twogap}, which reported an electron–phonon
coupling strength approximately five times larger than that obtained
from standard DFT. The origin of this discrepancy can be traced to the
different Fermi surface topologies predicted by the two approaches. In
both standard DFT and DFT+$U$, the Fermi surface is dominated by a large
Ni-$d_{x^2-y^2}$-derived sheet, accompanied by a small electron pocket
at the zone corner, with only minor modifications introduced by
Hubbard $U$. Consequently, the available scattering phase space is
only weakly affected, leading to a modest change in electron–phonon
coupling. In contrast, GW calculations predict a substantial
reconstruction of the Fermi surface, including a reversal of band
ordering and the emergence of a La-$d$-derived sheet at the Fermi
level. This qualitative change in Fermi surface topology significantly
enhances the electron–phonon coupling through the electronic
channel. Importantly, recent angle-resolved photoemission spectroscopy
(ARPES) experiments show a Fermi surface consisting of a large
Ni-$d_{x^2-y^2}$-derived sheet and a zone-corner electron pocket, in
good agreement with DFT results (see Supplementary Note
6 and Fig. S14 therein)~\cite{sciadv.adr5116}.

In contrast, RuO$_2$ exhibits a qualitatively different
behavior. While the electronic structure is moderately modified (see
Fig.~\ref{fig:fig6} and Fig. S15 in Supplementary Note 7), correlation effects
strongly renormalize the phonon spectrum by eliminating soft or
unstable modes and inducing significant phonon hardening. This leads
to a substantial suppression of electron–phonon coupling, primarily
through the phonon channel. As a result, the predicted superconducting
transition temperature is notably reduced and brought into better
agreement with experimental observations. While the ``partially
$U$-corrected'' approach is computationally inexpensive, our results
demonstrate that correlation effects can also have a profound impact
on both the phonon spectrum and the electron–phonon coupling
matrix. In the extreme case of TiO$_2$-strained RuO$_2$, standard DFT
predicts imaginary phonon modes, indicating a spurious lattice
instability. To stabilize the experimentally observed Pnnm structure,
Hubbard $U$ corrections are essential in phonon calculations. However,
if one combines the DFT phonon spectrum and electron–phonon matrix
with only partial $U$-corrections, the resulting electron–phonon
coupling is severely overestimated, yielding an unphysically large
coupling strength that is inconsistent with the low superconducting
transition temperature ($\sim 1.5$
K)~\cite{PhysRevLett.125.147001,Strain-stabilized}. Our results
therefore highlight that a self-consistent and reliable description of
electron–phonon interactions requires correlation effects---whether
treated via Hubbard $U$, GW, or hybrid functionals---to be
consistently incorporated not only in the electronic structure, but
also in the phonon spectrum and the electron–phonon coupling matrix.

Furthermore, for RuO$_2$ we find that while Hubbard $U$ generally
hardens phonons, anisotropic strain imposed by the TiO$_2$ substrate
can soften specific modes, such as those at Z point, consistent with
previous studies~\cite{PhysRevLett.125.147001,Strain-stabilized}.
This provides a possible explanation for why superconductivity emerges
in TiO$_2$-strained RuO$_2$ thin films but is absent in bulk. In
contrast, bulk RuO$_2$ is dynamically stable in the orthorhombic
$Pnnm$ phase, and standard DFT predicts a relatively large total
electron–phonon coupling ($\lambda \simeq 0.8$), which would suggest
an observable superconducting transition temperature. The experimental
absence of superconductivity in bulk RuO$_2$ therefore indicates that
correlation effects must be included to substantially suppress the
total electron–phonon coupling.

\section{Conclusion}

In summary, we implement an algorithm that integrates DFT+$U$ and finite displacement methods for phonon and electron-phonon $g$ matrix calculations. We apply this ``fully Hubbard $U$ corrected'' approach to two representative examples of infinite layer nickelates and ruthenium dioxide. For infinite-layer nickelates 20\% hole-doped LaNiO$_2$, regardless of hole doping, the Hubbard $U$ corrections slightly increase the electron-phonon interactions due to the additionally induced zone-center electron pocket. However, the total electron-phonon coupling of infinite-layer nickelates remains small by Hubbard $U$ corrections, contrasting with the full GW corrections~\cite{stevein_twogap} which substantially increase the DFT-calculated electron-phonon coupling by five times. We attribute this discrepancy to the differences in the Fermi surface topology between DFT+$U$ and GW methods. For TiO$_2$-strained RuO$_2$, the inclusion of Hubbard $U$ corrections is essential to eliminate imaginary phonon modes and stabilize the experimentally observed $Pnnm$ structure. Furthermore, the Hubbard $U$ corrections significantly reduce the total electron-phonon coupling of RuO$_2$, which alleviates the discrepancy between the large electron–phonon coupling from standard DFT calculations and the low experimental superconducting transition temperature of 1.5 K. Our work provides an approach to fully take into account the Hubbard $U$ corrections in electron-phonon interactions, which is applicable to a wide range of correlated materials.

\section{Methods}
We perform the density functional theory (DFT)
calculations~\cite{PhysRev.140.A1133,PhysRev.136.B864} using the
OpenMX package~\cite{PhysRevB.69.195113,PhysRevB.67.155108}. We use
the finite-displacement method to calculate the phonon spectra and
electron-phonon $g$ matrix~\cite{G.J.Ackland_1997,10.1063.5.0085759}. We employ
norm-conserving pseudopotentials and the local density approximation
(LDA)~\cite{PhysRevB.23.5048} for the exchange-correlation
functional. Hubbard $U$ corrections are applied to the Ni-$d$ and
Ru-$d$ orbitals to account for electron correlation effects, which are
consistently included in the calculations of the electronic structure,
phonon spectrum, and electron-phonon $g$ matrix. We consider a wide
and physically reasonable range of Hubbard $U$ value: $0-7$ eV for
infinite-layer nickelates and $0-5$ eV for ruthenium dioxide
(RuO$_2$).

For infinite-layer nickelates, the hole doping is modelled by
changing the total number of electrons~\cite{PhysRev.157.544}. For
ruthenium dioxide, we first fully relax the crystal structure of bulk
RuO$_2$. Then in order to simulate the epitaxial strain imposed by
TiO$_2$ substrate, we increase the theoretical $[1\overline{1}0]$
lattice constant of RuO$_2$ by 2.3\% and reduce the $[001]$
theoretical lattice constant by 3.1\%, consistent with the
experiment~\cite{PhysRevLett.133.176401,PhysRevLett.125.147001}. With
those two fixed lattice constant, we fully relax the internal atomic
coordinates and the [110] lattice constant.

We use an energy cutoff of 125 Hartree and adopt La8.0-s3p2d2f1,
Ni6.0H-s3p2d1, O6.0-s2p2d1 and Ru7.0-s3p2d2 pseudo-atomic orbitals,
which contain 26, 14, 13 and 19 basis functions in each atom. The
electronic Brillouin zone is sampled with a $12\times 12 \times 12$
Monkhorst-Pack $\mathbf{k}$ mesh for both infinite-layer nickelate and
ruthenium dioxide. Self-consistency in the electronic structure
calculations is converged to $1\times10^{-9}$ Hartree. Both cell and
internal atomic positions are fully relaxed until each force component
is smaller than $1\times10^{-6}$ Hartree/Bohr. In the finite
displacement calculations for phonon spectrum, we enlarge the original
simulation cell to a supercell. The size of the supercell is
$5\times5\times5$ for infinite-layer nickelate and $3\times3\times3$
for ruthenium dioxide.  The \textbf{k} mesh for the supercell is
$2\times2\times2$ for infinite-layer nickelate and $4\times4\times4$
for ruthenium dioxide. In the electron-phonon calculations, we use the
same \textbf{k}-grid and \textbf{q}-grid of $40 \times 40 \times
40$. The smearing width of electronic structure calculations is 50 meV
and the smearing width of phonon calculations is 0.5 meV. The
temperature is set to 20 K. In electron-phonon calculations, we only
take into account the bands close to the Fermi level, since only those
bands make contributions to the total electron-phonon coupling. For
infinite-layer nickelates, the number of bands we consider in the
electron-phonon calculations is 2; for ruthenium dioxide, the number
of bands we consider is 4.

With the total electron-phonon coupling $\lambda$, we can use the
McMillan-Allen-Dynes formula~\cite{Tc1,Tc2} to estimate the
superconducting transition temperature $T_c$:
\begin{equation}
  \label{eqn:MAD}
 T_c=\frac{\omega_{\textrm{log}}}{1.2}\textrm{exp}\left[\frac{-1.04(1+\lambda)}{\lambda(1-0.62\mu^{*})-\mu^{*}}\right]    
\end{equation}
where $\omega_{\textrm{log}}$ is a logarithmic average of the phonon frequency~\cite{Tc2,Preempted}, defined as:
\begin{equation}
 \omega_{\textrm{log}}=\textrm{exp}\left[\frac{2}{\lambda}\int \frac{d\omega}{\omega}\alpha^2F(\omega)\textrm{log}\omega \right]
\end{equation}
$\mu^{*}$ is the Anderson-Morel parameter that describes the screened Coulomb
interaction and is treated as an adjustable parameter. Since we
already take into account Hubbard $U$ corrections in our calculation
of electron-phonon interactions, we set $\mu^{*}=0$. 

\section*{Acknowledgement}
We are grateful to Qijing Zheng and Jinjian Zhou for useful discussions. H.C was financially supported by the National Natural Science Foundation of China under project number 12374064 and 12434002, and Science and Technology Commission of Shanghai Municipality under grant number 23ZR1445400. J.Z. was supported by the National Natural Science Foundation of China under project number 12125408 and 12334004. C.X. was supported by the National Natural Science Foundation of China under project number 12404082. NYU High-Performance-Computing (HPC) provides computational resources.

\section*{AUTHOR CONTRIBUTIONS}
H.C. and J.Z. conceived and supervised the project. J.C. and Y.T. implemented the code and performed the calculations. C.X. participated in the analysis. J.C. and H.C. wrote the manuscript. All the authors participated in the discussion and the preparation of the manuscript.

\section*{DATA AVAILABILITY}
The data supporting the conclusions of this study are provided in the Supplementary Information. Any additional data relevant to the study's findings are available from the corresponding author upon reasonable request.


%

\end{document}